\documentclass[useAMS,usenatbib]{mn2e}
\usepackage{graphicx}
\newcommand {\be}{\begin{equation}} 
\newcommand{\ee}{\end{equation}}    
\newcommand{\sss}{\scriptscriptstyle}
\def\nabp{\nabla_{\perp}}

\def\vti{v_{{\sss T}i}}

\def\vte{v_{{\sss T}e}}
\def\vtj{v_{{\sss T}j}}

\title[Drift waves in the corona: heating and acceleration of ions at frequencies far below the gyro frequency]{Drift
waves in the corona: heating and acceleration of ions at frequencies far below the gyro frequency}
\author[J. Vranjes and S. Poedts]
{J. Vranjes\thanks{E-mail: Jovo.Vranjes@wis.kuleuven.be; jvranjes@yahoo.com} and S. Poedts\thanks{E-mail:
Stefaan.Poedts@wis.kuleuven.be }
\\
K.U.Leuven, Center for Plasma Astrophysics, Celestijnenlaan 200B, 3001 Leuven,
 Belgium,\\ and Leuven Mathematical Modeling and Computational Science Center
 (LMCC)
}
\begin{document}

\date{Accepted xxx. Received xxx; in original form xxx}

\pagerange{\pageref{firstpage}--\pageref{lastpage}} \pubyear{2010}

\maketitle

\label{firstpage}

\begin{abstract}
In the solar corona, several mechanisms of the drift wave instability can make the mode growing up to amplitudes at
which particle acceleration and stochastic heating by the drift wave take place.  The stochastic  heating, well known
from laboratory plasma physics where it has been confirmed in numerous experiments, has been completely ignored in past
studies of coronal heating. However, in the present study and in our very recent works it has been shown that the
inhomogeneous coronal plasma is, in fact, a perfect environment for fast growing drift waves. As a matter of fact, the
large growth rates are typically of the same order as the {\bf wave} frequency. The consequent heating rates may exceed the
required values for a sustained coronal heating by several orders of magnitude. Some aspects of these phenomena are
investigated here. In particular the analysis of the particle dynamics  within the growing wave is compared with the
corresponding fluid analysis. While both of them predict  the stochastic heating, the threshold for the heating
obtained from the single particle analysis is higher. The explanation for this effect is given.

\end{abstract}

\begin{keywords}
Sun: corona, oscillations.
\end{keywords}

\section{Introduction}

In our recent works \citet{v1,v2,v3, v4} the drift wave theory has been applied to the problem of the heating of the
solar corona. The drift wave is driven by gradients of the background plasma parameters (e.g., gradients of the plasma
density, temperature, and magnetic field) that may act either separately or all together. In the literature, the drift
wave is frequently called the {\em universally unstable  mode} because it is growing both within the (two- or
multi-component) fluid theory and the kinetic theory. In the former (fluid) description, it is either a collisional
instability (discussed in \citet{v5} in the context of the solar atmosphere) that appears as a combined effect of the
electron collisions, the ion inertia, and the density gradient, or a reactive one \citep{v4} in the presence of all
three gradients mentioned above where now the ions play a major role. The latter (kinetic) theory, on the other hand,
can also describe these two instabilities. However, the kinetic description yields an additional (purely kinetic)
instability that is absent within the fluid description. In application to the solar corona, this additional kinetic
instability is studied in \citet{v1,v2}. This purely kinetic growth of the drift wave is entirely due to electron
dynamics and it develops provided the wave frequencies are below the electron diamagnetic drift frequency.

Regardless which of the mentioned instabilities is in action, the result is always a growing wave amplitude. When some
critical value for the wave potential is achieved, stochastic heating sets in \citep{mc,mc2,san}. By  nature  this
heating is  strongly anisotropic (acting mainly in the direction perpendicular to the magnetic field vector). Moreover,
it is mass dependent (heating mainly ions, with better heating of the heavier ions), it is due to the electrostatic
features of the drift wave, and it is a low frequency process (typically the drift wave frequency is much below the ion
gyro-frequency).

The drift wave can easily become electromagnetic in case of a relatively high plasma-$\beta$ (exceeding the electron to
ion mass ratio). In that limit, it is in fact coupled to the Alfv\'{e}n wave. Yet, even in this case the heating
develops almost without change and primarily due to the electric field of the drift-Alfv\'{e}n mode. In the past, this
has been studied both analytically and numerically, and it was even experimentally verified in laboratory (tokamak)
plasmas \citep{mc,mc2,san,white}. For the coronal environment the electromagnetic  drift-Alfv\'{e}n case has been
discussed in detail  in \citet{v6}.

For the purpose of the present study, in order to quantitatively describe some details of the heating and acceleration
of plasma particles, any of the mentioned instabilities of the drift wave will do. Hence, we shall use the kinetic,
density gradient driven instability from \citet{v1,v2}.


\section{The instability of the density gradient driven drift wave}

From standard textbooks \citep{ichi,weil}, the frequency and the growth rate of the kinetic drift wave are given by
\be
\omega_r=-\frac{\omega_{*i} \Lambda_0(b_i)}{1- \Lambda_0(b_i) + T_i/T_e + k_y^2 \lambda_{di}^2},\label{k1} \ee
and
\[
\gamma \simeq -\left(\frac{\pi}{2}\right)^{1/2} \frac{\omega_r^2} {|\omega_{*i}| \Lambda_0(b_i)}\left[\frac{T_i}{T_e}
\frac{\omega_r - \omega_{*e}}{|k_z|\vte} \exp[-\omega_r^2/(k_z^2 \vte^2)] \right.
\]
\be
\left.
 +
\frac{\omega_r - \omega_{*i}}{|k_z|\vti} \exp[-\omega_r^2/(k_z^2 \vti^2)]\right]. \label{k2} \ee
Here, the geometry is chosen as follows. The constant magnetic field is given by $\vec B_0=B_0 \vec e_z$, the
equilibrium density gradient in the electrically neutral plasma with $n_{i0}=n_{e0}=n_0$ is given by $\nabp n_0= - \vec
e_x n_0'\equiv -\vec e_x d n_0/dx$, and the perturbations are assumed to be of the form $f(x) \exp(- i \omega t + i k_y
y + i k_z z)$, with $|k_z|\ll |k_y|$ and an $x$-dependent wave amplitude $f$. The local approximation will be used.
Hence, it will be  assumed that $|d/dx|\ll |k_y|$.

In the derivations of Eqs.~(\ref{k1}) and {\ref{k2}),  the condition of a  strongly magnetized plasma is used,
$|\omega|\ll \Omega_i$, together with the smallness of the acoustic response of ions in the direction parallel to the
magnetic field vector, implying that $\omega_r/k_z\gg c_s$, where $c_s$ is the ion sound speed. The notation in
Eqs.~(\ref{k1}) and (\ref{k2}) is standard \citep{weil}, i.e.,
\[
 \omega_{*e}=-k_y \frac{\vte^2}{\Omega_e} \frac{n_{e0}'}{n_{e0}}, \quad  \omega_{*i}=- (T_i/T_e)\omega_{*e}, \quad
 \vtj^2=\kappa T_j/m_j,
 \]
\[
 \Lambda_0(b_i)=I_0(b_i) \exp(-b_i), \quad b_i=k_y^2 \rho_i^2,  \quad \lambda_{di}=\vti/\omega_{pi},
\]
where $I_0$ is the modified Bessel function of the first kind and of the order 0.

From Eq.~(\ref{k2}) it is seen that within the present instability, the  sign of $\gamma$ can be changed only by the
electron term if $|\omega_r|<|\omega_{*e}|$. This is, however, only a necessary condition. A sufficient condition is
obtained when the contribution of  the rest of the expression (the ion damping terms) is taken into account. As
demonstrated by various graphs in \citet{v1,v2}, in the case of the solar corona it is more difficult to find a regime
where the mode is damped than a regime where it is growing. In  these studies it is  shown that the mode grows more
strongly for shorter perpendicular wave lengths $\lambda_y$ (meter or sub-meter size) and for longer parallel
wavelengths $\lambda_z$ that may exceed the perpendicular ones for 4 to 5 orders of magnitude. Note that a similar
ratio of the two components $\lambda_j$ holds for tokamak plasmas as well.  The same  conclusions can also be drawn in
the case of the reactive (the so-called
 $\eta_i$-instability) discussed in \citet{v4}.

To provide some details of the particle acceleration and heating by the growing drift wave in the context of the solar
corona, we need  specific data. Hence,  as an example we here use a set of parameters  from \citet{v1,v2} that are
known to yield the instability: $B_0=10^{-2}\;$T, $n_0=10^{15}\;$m$^{-3}$, $L_n= [(dn_0/dx)/n_0]^{-1}=s\cdot 100\;$m,
$\lambda_y=0.5\;$m, and $\lambda_z=s \cdot 10^4\;$m. These are just representative parameters with no particular
significance. As discussed in \citet{v2}, the instability will in fact develop even when the density is varied by
several orders of magnitude around the given value, and the same holds for the magnetic field and the temperature. The
parameter $s$ is introduced in \citet{v1,v2} for convenience only, because it was realized that the ratio
$\gamma/\omega_r$ remains exactly the same in case the ratio $\lambda_z/L_n$ is fixed. Here,  $L_n$ is the
characteristic scale-length for the inhomogeneous equilibrium density. This further implies that we may go to very
different scales, e.g., by taking $s$ in the range $0.1 - 10^4$. As long as $\lambda_y$ is kept constant, the ratio
$\gamma/\omega_r$ will remain the same, although the actual  values for these two quantities will certainly change.
Hence, for the given parameters, from Eqs.~(\ref{k1}) and (\ref{k2}) one finds $\gamma/\omega_r=0.26$. In the case
$s=1$, this in fact implies $\omega_r=254\;$Hz and $\gamma=66\;$Hz. While in the case $s=10^3$, this would give a mode
with $\omega_r=0.254\;$Hz and $\gamma=0.66\;$Hz. Going to such small values of $L_n$ (tens of meters) in fact makes
sense because the perpendicular diffusion, that naturally develops in such a plasma with  density gradients, is indeed
very small, with the diffusion coefficient $D_\bot$ of the order of $0.01\;$m$^2$/s and the corresponding diffusion
velocity $D_\bot \nabla_\bot n/n$  of just a few millimeters  per second \citep{vx}. Short values of $L_n$ in the same
time imply very high frequency drift waves and fast growing instabilities, so that the expected heating will develop at
time scales that are orders of magnitude shorter than the plasma diffusion. Note that the parameters $\lambda_z$ and
$L_n$ can of course also be varied independently of each other.

\section{Physical mechanism of heating by the drift wave}

\subsection{Heating within the collective interaction frame}

The physics of the mentioned stochastic heating is described in detail in various  sources. Here, we give some short
sketches of it by  following \citet{bel}, and using the standard drift wave theory. Within the two-fluid theory, the
perpendicular velocity of the ion unit volume   is given by the recurrent formula \citep{v5, v2}
\[
v_{i\bot}=\frac{1}{B_0} \vec e_z \times \nabp \phi
 + \frac{v_{{\sss T}
i}^2}{\Omega_i} \vec e_z \times \frac{\nabp n_i}{n_i} + \vec e_z \times \frac{\nabp\cdot \pi_i}{m_i n_i \Omega_i}
\]
\be
  + \frac{1}{\Omega_i} \frac{d}{dt} \vec e_z \times \vec v_{i \bot}, \quad \frac{d}{dt}\equiv  \frac{\partial}{\partial t} + \vec v\cdot\nabla.
\label{e3} \ee
The four terms here correspond to, respectively, the  $\vec E\times \vec B$-drift that is usually the leading order
one, the diamagnetic drift $\vec v_{*i}$, the stress tensor drift $ \vec v_{\pi i}$, and the polarization drift $\vec
v_{pi}$. The velocity  can be calculated up to small terms of any order using the drift approximation
$|\partial/\partial t|\ll \Omega_i$. In some situations, in plasmas with relatively cold ions,  as a first
approximation one may assume the  $\vec E\times \vec B$-drift as the leading order one, and set it   into the velocity
$\vec v_{i \bot}$ in the polarization drift. The diamagnetic drift does not contribute to the ion flux $\nabla\cdot(n
\vec v_{*i})\equiv 0$, and also its contribution to the convective derivative of the polarization drift is canceled by
the part of the stress tensor flux $n \vec v_{\pi i}$ \citep{vmm}.  Therefore, we shall focus now on the $\vec E\times
\vec B$-drift, and the polarization drift $\vec v_{pi}=(d \vec E_\bot/dt)/(B_0 \Omega_i)$, where $\vec E_\bot\equiv -
\vec e_y k_y \phi_1$. These two  together describe an elliptical motion in the presence of the wave, described  by:
 \be
\vec v_{\sss E}=\vec e_x \frac{k_y \phi_1}{B_0}  \sin(k_y y + k_z z - \omega_r t), \label{e4}
 \ee
and
\[
 \vec v_{pi}=-\vec e_y \frac{\omega_r k_y \phi_1(t)}{B_0 \Omega_i}\left[\left(1- \frac{k_z}{\omega_r} \frac{d z}{dt}\right)
  \cos \varphi-\frac{\gamma}{\omega_r}\sin \varphi\right]\times
\]
\be
 \times 1/ \left(1-\frac{k_y^2 \phi_1(t)}{B_0 \Omega_i} \cos \varphi\right), \quad \varphi= k_y y + k_z z - \omega_r t.
\label{e5}
 \ee
In deriving Eq.~(\ref{e5}) the wave amplitude is taken as time dependent due to the wave instability,
$\phi_1=\widehat{\phi}\exp(\gamma t)$, $\widehat{\phi}=\;$const.,  and the same holds for the coordinates $y(t), z(t)$
appearing in the wave phase $\varphi(t)$. That is the reason for the appearance of the terms with $\gamma$ and $d
z/dt$. Neglecting these two terms, and taking a constant potential, Eq.~(\ref{e5}) becomes identical to the
corresponding expression from \citet{bel}.

Crucial for the heating is the term $[k_y^2 \phi_1(t)/(B_0 \Omega_i)] \cos \varphi$ in the denominator. Looking back
into the derivations, it turns out that it appears from the derivative $d (\sin \varphi)/dt\rightarrow d
y/dt=v_y=v_{pi}$. Keeping it here  has sense only in case of a relatively large displacement in the direction of the
perpendicular wave vector. From Eq.~(\ref{e5}) it is seen that for
\be
a(t)\equiv k_y^2 \phi_1(t)/(B_0 \Omega_i)=k_y^2 \rho_i^2 e \phi_1(t)/(\kappa T_i)\geq 1, \label{a} \ee
 the denominator may vanish and, consequently,  the polarization drift should go to infinity. Note however, that Eq.~(\ref{e5})
is obtained approximately starting from the guiding center approximation that becomes
 violated for such a large polarization drift, and the analytical expression for the polarization drift in this limit becomes  inaccurate.
Yet, Eq.~(\ref{e5}) describes  the general trend for  the polarization drift;   a fast and significant acceleration and
heating of the plasma particles will definitely take place, as experimentally verified in \citet{mc,mc2,san}.

\subsection{Individual particle dynamics}

As discussed in \citet{bel}, a direct numerical integration for plasma particles is required for a full validation of
the above described heating mechanism for the solar corona. The numerical simulation has to take into account the
actual values of the electric field in the position of a specific gyrating plasma particle.  For that purpose, taking
again the drift-wave electric field with a time-varying amplitude, the appropriate set of equations for a single
particle reads
\be y''(t) + y(t)- a(t) \sin\left[y(t) + \frac{k_z}{k_y} z(t) - bt\right]=0, \label{e6} \ee
\be z''(t) - a(t) \frac{k_z}{k_y} \sin\left[y(t) + \frac{k_z}{k_y} z(t) - bt\right]=0, \label{e7} \ee
\be
x'(t)-y(t)=0. \label{e8} \ee
Here, the prime denotes a derivative in time, and  the spatial and time  coordinates are normalized as $x, y, z
\rightarrow k_y x, k_y y, k_y z$, $t\rightarrow \Omega_i t$, $b=\omega_r/\Omega_i$. Equation~(\ref{e8}) is obtained
after one integration where an integration constant appears that has been neglected. Note that Eq.~(\ref{e6}) can be
reduced to the Mathieu equation with a source term. Namely, neglecting the $z$-terms and after introducing $\tau=
bt/2$, for the small remaining argument $\varphi$ it can be written in the generic form
\be
d^2 y/d\tau^2 + [\alpha - 2 q(t) \cos 2\tau] y(\tau) =  c(\tau) \sin 2 \tau, \label{m} \ee
where $\alpha=4/b^2$, $q(\tau)=2 \alpha(\tau)/b^2$, and $c(\tau)=- 4 \alpha(\tau)/b^2$. The solutions of this equation
without the right-hand side are the Mathieu functions $M(\alpha, q, \tau)$. Equation~(\ref{m}) has unstable solutions
and it is discussed in detail in {\bf \citet{chen},} \citet{white}.

\begin{figure}
\includegraphics[height=6cm, bb=15 15 285 225, clip=,width=.95\columnwidth]{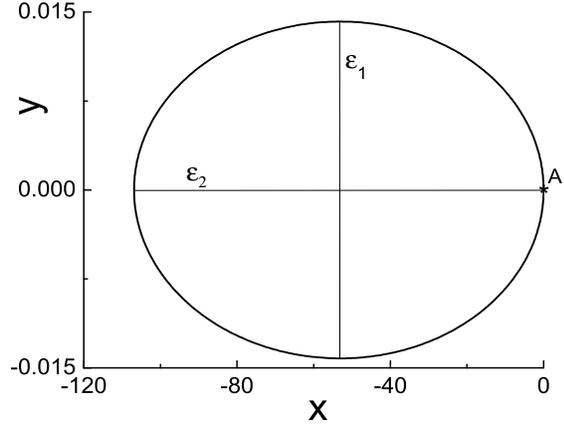}
 \caption{Particle positions in the $x,y$-plane within a wave period, after it starts from  the point A with $(x, y, z)=(0, 0, 0)$, for a small  wave amplitude $\phi_1=0.86$ V.}\label{fig1}
\end{figure}

\begin{figure}
\includegraphics[height=6cm, bb=15 15 285 225, clip=,width=.95\columnwidth]{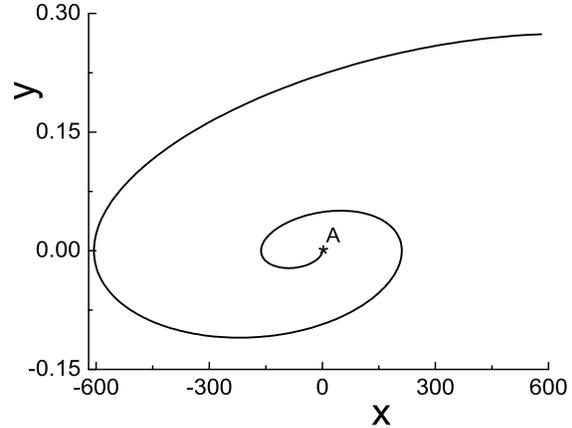}
 \caption{Normalized particle positions in the $x,y$-plane after it starts from the point A with $(x, y, z)=(0, 0, 0)$,
  for a growing electric field potential $\phi_1(t)=\widehat{\phi}\exp(\gamma t/\Omega_i)$, and  $\widehat{\phi}=0.86$ V.}\label{fig2}
\end{figure}

The set of ordinary equations (\ref{e6})-(\ref{e8}) is solved numerically. We do this first for a very small and
constant wave amplitude in order to demonstrate some basic features of the particle motion due to the wave electric
field. The result is  shown in Fig.~1, where we give the projection of the proton positions in the perpendicular plane,
$x(t), y(t)$,  in the wave field with the constant amplitude $e \phi_1/(\kappa T_i)=0.01$, that corresponds to
$\phi_1=0.86\;$V. Here and further in the text we take  $\lambda_y=0.5$ m, $\lambda_z= 2\cdot 10^4$ m,  which
consequently from Eq. (\ref{k1}) yields the wave frequency $\omega_r=254$ Hz. The  normalization  is the same as above,
viz.\ $(x, y)\equiv (k_y x, k_y y)$. This all yields $a=0.014$, hence the stochastic heating is absent, and the
particle trajectory is a very elongated ellipse, with the motion almost completely in the $x$-direction  due to the
$\vec E\times \vec B$-drift described above (observe the difference in $x,y$-scales). After a wave period the particle
returns to the starting position~A  with the coordinates $(x, y)=(0, 0)$. The ratio of the two axes of the ellipse
$\varepsilon_1/\varepsilon_2=0.00013$ describes the fact that the particle motion is mainly due to the $\vec E\times
\vec B$-drift in the $x$-direction, and that the polarization drift in the direction of the wave-vector (the
$y$-direction) is negligible.

Performing the same plot for electrons yields the same $\epsilon_2$ as expected (the $\vec E\times \vec B$-drift is
 independent on mass), while $\epsilon_1$ becomes reduced by the factor $m_e/m_i$ (the polarization drift separates charges and it  is mass
dependent).

\begin{figure}
\includegraphics[height=6cm, bb=15 15 285 225, clip=,width=.95\columnwidth]{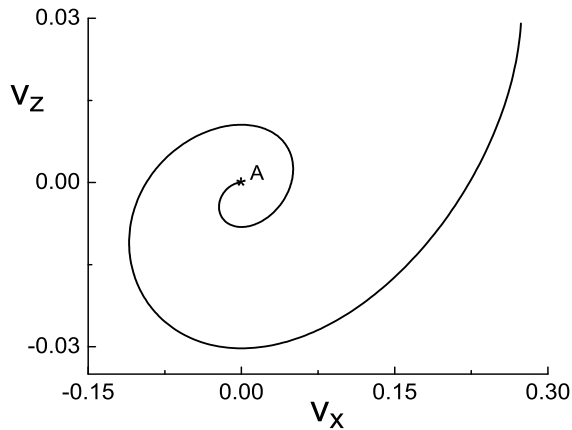}
 \caption{Normalized velocities $v_x, v_z$ for the same parameters as in Fig.~2.}\label{fig3}
\end{figure}

\begin{figure}
\includegraphics[height=6cm, bb=15 15 285 225, clip=,width=.95\columnwidth]{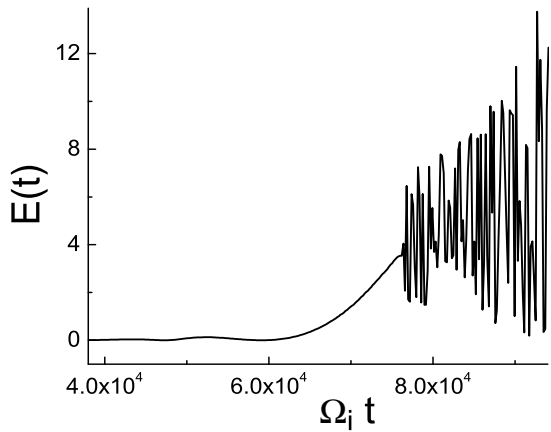}
 \caption{Change in time of the normalized kinetic energy of a particle with a unit mass $E=(v_x^2+v_y^2+ v_z^2)/2$.}\label{fig4}
\end{figure}

On the other hand, taking a time-varying potential $\phi_1(t)$ due to the above demonstrated drift wave instability,
and restricting the time interval to relatively short values (below the onset time of the stochastic heating), yields a
spiral trajectory in the $x, y$-plane. One example of this is presented in Fig.~2. Here, the starting value for the
potential is the same as above $e \phi_1(0)/(\kappa T_i)=0.01$, and the same holds for the particle position. The other
parameters are the same as in Fig.~1. With the same normalization  the potential is given by
$\phi_1(t)=\widehat{\phi}\exp(\gamma t/\Omega_i)$, $\gamma=0.26 \omega_r$, $\omega_r=254\;$Hz,
$\widehat{\phi}=\phi_1(0)$. The maximum time in physical units here is $0.04\;$s, and in this moment the potential has
reached the value of $17\;$V. Comparing with Fig.~1, one observes that the displacement in the $y$-direction due to the
polarization drift is now increased by a factor 20.

It is interesting to compare the leading perpendicular $\vec E\times \vec B$-drift velocity $v_x$, and the parallel
velocity $v_z$ (along the magnetic field vector). The parametric plot in Fig.~3 shows that for these potential
amplitudes the parallel velocity $v_z$ remains about one order of magnitude smaller.

In order to  see what happens for larger amplitudes of the electrostatic drift wave when the  stochastic heating is
supposed to be in action, Eqs.~(\ref{e6})-(\ref{e8}) are solved by allowing a slightly larger time range. One of the
results is shown in Fig.~4 for the kinetic energy of a particle with   unit mass $E(t)/m=[v_x(t)^2+v_y(t)^2 +
v_z(t)^2]/2$, with normalized velocities as above. The other parameters are the same as in the previous text and
figures. Here, the stochastic heating takes place after around $0.078\;$s, in the moment when the growing wave
amplitude reaches the value of around $150\;$V. Note that this is by about a factor $2.5$ larger than the value
obtained from the condition (\ref{a}) which, in fact, follows  from an approximative procedure as explained earlier.
Thus, a higher necessary threshold for the stochastic heating is expected.  However, a  plasma can support multiple
waves in the same time.
 The drift wave spectrum described by
Eqs. (\ref{k1}, \ref{k2}) in realistic situations imply the presence of more waves rather than a single one.
The analysis presented in a recent study \citep{sheng} shows that in such cases the instability threshold can be
considerably reduced. Therefore, the ion heating by the mechanism which we discuss here will  be even more efficient.

\begin{figure}
\includegraphics[height=6cm, bb=15 15 270 225, clip=,width=.95\columnwidth]{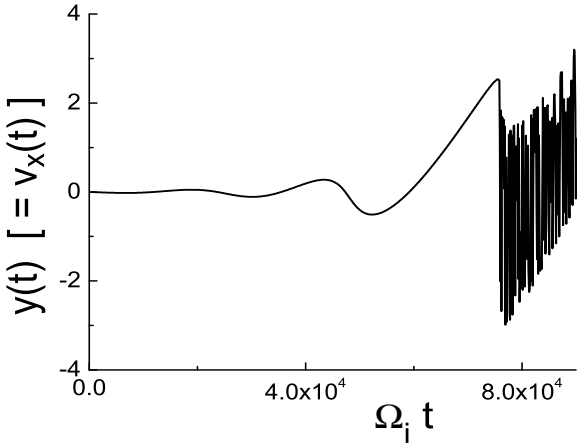}
 \caption{Normalized displacement in the direction of the polarization drift $y(t)$, and  the  normalized perpendicular velocity component
  $ y(t) \equiv v_x(t)$.
}\label{fig5}
\end{figure}

\begin{figure}
\includegraphics[height=6cm, bb=15 15 270 225, clip=,width=.95\columnwidth]{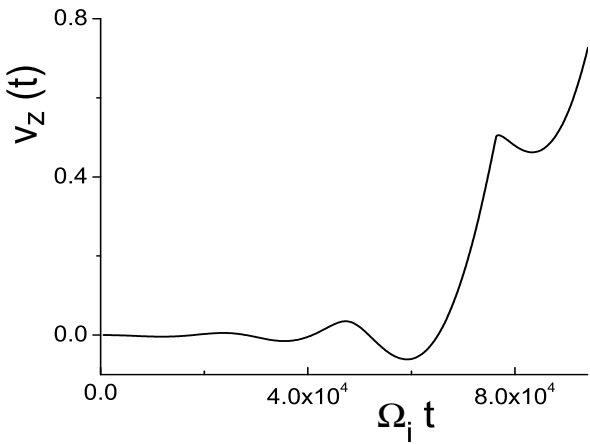}
 \caption{Normalized perturbed velocity  in the direction parallel to the magnetic field.
}\label{fig6}
\end{figure}

The corresponding plot of the displacement in the direction of the wave $y(t)$, which is in fact also  equal to the
velocity $v_x(t)$ [see Eq.~(\ref{e8})]   is shown in Fig.~5. The plot of the velocity $v_y(t)$ (that is not given here)
reveals  the  amplitude that is completely negligible until $\Omega_i t$ becomes close to $8\cdot 10^4$; after that it
is very stochastic and with the amplitude equal to that of the velocity $v_x(t)$. Thus  we are in the range of
parameters in which the expansion procedure used previously is not valid any longer because the polarization and $\vec
E\times \vec B$ drifts are of the same order, and the stochastic heating is fully in action. Note that, although the
motion is stochastic/chaotic, it is in fact
 deterministic and not random as commented in \citet{bel}.  Observe that the total physical time interval for Figs.~4 and 5 is only about 3-4 wave-periods.

The plot of the parallel velocity $v_z(t)$, presented in Fig.~6, shows that along the magnetic field the particle
dynamics remains non-stochastic, as should be expected in view of the analysis presented in the text above. The
velocity magnitude is  below  1, thus  still smaller than the perpendicular velocity (see Fig.~5) which is the
consequence of the stochastic heating.

The theory presented in \citet{mc,mc2,san} predicts the maximum stochastic velocity
 $ v_{max}\simeq [k_y^2 \rho_i^2 e \phi/(\kappa T_i) + 1.9]\Omega_i/k_y$, which  further yields the effective achieved stochastic temperature
 $T_{max}=m v_{max}^2/(3 \kappa)$. For the parameters used above this yields  $T_{max}\simeq 2\cdot 10^6\;$K, i.e., the
 increase of the temperature for one million Kelvin within the growth time $\tau_g\simeq 0.1\;$s. The maximum energy released by this mechanism is $\Sigma_{max}= n_0 m_i
 v_{max}^2=0.04\;$J/m$^{-3}$. The energy release rate, $ \Sigma_{max}/\tau_g$,  exceeds for  several orders of magnitude \citep{v1, v2} the value presently accepted by researchers as required
 for a sustained coronal heating.

\section{Summary}
The present study provides some additional details of our recently proposed model for the coronal heating, which
implies the heating by electrostatic drift waves.     The heating by waves, which presently represents  one of two
major heating scenarios,   has been studied in many works from  the early days of the coronal heating investigation,
e.g., by  \citet{alf} and  \citet{pid},  but  also more recently in \citet{pek, suz}. More details about the existing
heating models are available in various sources \citep{nar, klim}, some of them have also been  discussed in more
detail in our recent work \citet{v2}.

 In this work, the trajectories of individual plasma particles have been investigated within the framework of the
drift wave instability and the associated stochastic heating, with the aim to describe some fine details of the
particle dynamics during  the heating process. The calculation of the actual growth rates of the mode is not repeated
here because it has been given in detail in our recent studies. From the present study and from our recent works it
follows that the general trend of the ion behavior in the wave field with a growing amplitude can be described
analytically from the fluid point of view, as discussed here in Sec.~3.1. However, a more detailed picture is obtained
by following a single particle which is moving in the wave field. For example, this is seen from Sec.~3.2 in the case
of the critical threshold for the onset of the heating. We observe that the general theory of the drift wave
instability presented in some early studies \citep{ter}, and  not necessarily related to the heating, also includes the
particle aspect approach, and as such, is in perfect agreement with the fluid and  kinetic modeling.

It is also interesting to note that, although the individual particle dynamics becomes  stochastic for relatively large
wave amplitudes, the overall picture of the macroscopic motion of the plasma, caused by the wave, actually remains well
described by the fluid description. This has been directly experimentally verified by \citet{bail}. The parameter $a$
[c.f.\ Eq.~(\ref{a})] in their experiment was around 2, and thus the stochastic heating was in action, yet the observed
electrostatic potential and density profiles  nicely  matched the  theoretical contour plots corresponding  to a wave
described within the fluid theory and drift approximation. This is in agreement with the statement given in Sec.~3.2
about the nature of the heating; the increased effective temperature $T_{max}$ follows from the stochastic velocity $
v_{max}$ which in fact  is not random, yet in practical terms it is indistinguishable from standard collisional
thermalization and the broadening of the distribution function \citep{bel}.

The model used here implies a  maxwellian distribution function for plasma species, that is the basis for both the
kinetic and the fluid description of the instability. However, the stochastic heating with all its features \citep{v2,
v4} in fact implies that such a starting distribution function will necessarily evolve. The perpendicular ion
temperature will grow and eventually become larger than the parallel one. Hence, the particle distribution may pass
through several stages, going through the Oort's distribution with the temperature anisotropy and  to the more general
loss-cone distribution \citep{bar}, or to something else. This opens an interesting possibility for eventual numerical
simulations because such fine details are beyond the scope of an analytical study.

\section*{Acknowledgments}

The  results presented here  are  obtained in the framework of the projects G.0304.07 (FWO-Vlaanderen), C~90347
(Prodex),  GOA/2009-009 (K.U.Leuven). Financial support by the European Commission through the SOLAIRE Network
(MTRN-CT-2006-035484) is gratefully acknowledged.

\bsp

\label{lastpage}

\end{document}